\documentclass{aastex631}

\begin{document}

\title{Possible detection of the progenitor of the Type II supernova SN2023ixf}

\correspondingauthor{Joanne Pledger}
\email{jpledger@uclan.ac.uk}

\author[0000-0001-8907-3051]{Joanne L Pledger}
\affiliation{Jeremiah Horrocks Institute, University of Central Lancashire, Preston, PR1 2HE, UK}

\author{Michael M Shara}
\affiliation{American Museum of Natural History, New York City, NY 10024, USA}



\begin{abstract}
Stellar evolution theory predicts multiple pathways to the explosive deaths of stars as supernovae. Locating and characterizing the progenitors of well-studied supernovae is important to constrain the theory, and to justify and design future surveys to improve on progenitor detections. Here we report the serendipitous pre-explosion imaging, by the {\it Hubble Space Telescope}, of SN2023ixf, one of the nearest extragalactic supernovae ever discovered, in the galaxy M101. The extremely red color and absolute magnitude M$_{F814W}$\,=\,--5.11$^{+0.65}_{-0.47}$\,mag suggest that the progenitor was a red supergiant. Comparison with stellar evolutionary isochrones suggests it is within the relatively low initial mass range of $\sim$8-10M$_{\odot}$ and that there is likely a lot of dust present at the supernova site.

\end{abstract}



\section{Introduction} \label{sec:intro}

{\bf Supernovae are responsible for the production of many of the chemical elements and dust in galaxies and throughout the universe \citep{Burbidge1957,Cameron1957,Hoyle1970,Johnson2020}, the heating of the interstellar medium  and galactic fountains \citep{McGee1975,Corbelli1988}, cosmic rays \citep{Zwicky1934,Blandford1978}, long-duration gamma-ray bursts \citep{Cano2017} and possibly even the formation of our own Solar System \citep{Clayton1982}. They are the most luminous distance indicators known, responsible for the detection of the dark energy \citep{Riess1998,Perlmutter1999}. Understanding how supernovae come about is thus important to almost every field in astrophysics. The most basic question that must be answered to make progress in understanding supernovae is: What stars become supernovae? }

Nearly 50 years ago Beatrice Tinsley \citep{Tinsley1975} posed that question, and summarized everything then known about the subject in an 11 page review article. In the interim over 5600 refereed journal articles have greatly advanced both observations and stellar evolution theory to yield some answers, but half a century later there are still multiple types of supernovae for which the answer remains ``We still don't know for certain". {\bf This ignorance is a major impediment to testing theories of the late stages of stellar evolution in both single and binary stars \citep{Smartt2009,Langer2012,Sukhbold2016,Williams2019,Rodriguez2022,liu2023}.}


\textbf{To date there have been $\sim$30 direct detections of Type II SN progenitors (see \citet{VanDyk2017} for a review), the first being SN1961V in NGC\,1058 \citep{Bertola1964, Zwicky1964} and the closest being SN1987A in the LMC \citep{Gilmozzi1987}. Many pre-SN observations only yield an upper limit; successful detections of progenitors are typically limited to $\sim$30\,Mpc}. At a distance of just 6.4\,Mpc \citep{Shappee2011} SN2023ixf in M101 is one of the closest and brightest extra-galactic supernovae ever detected. Intensive followup studies at many wavelengths are underway, \textbf{in the X-ray (ATel \#16049; \cite{Grefenstette2023} and ATel \#16044; \cite{Kawai2023}.), optical (ATel \#16045; \cite{Villafane2023}) and also a hunt for the associated neutrinos (ATel \#16043; \cite{Thwaites2023}}. This will be one of the most intensively studied supernovae ever, thus characterizing its progenitor is clearly of importance.

SN2023ixf in M101 was discovered by K. Itagaki on 19th May 2023 with a position of $\alpha$(J2000)\,=\,14:03:38.564 and $\delta$(J2000)\,=\,54:18:42.02, \citep{Itagaki2023} and was classified as a Type II SN shortly after by \citet{Perley2023}. In this paper we use archival {\it Hubble Space Telescope} ACS/WFC imaging to identify the possible progenitor of this event and use evolutionary models to predict an initial mass. \\

\section{Observations of the SN site \& Data Reduction} \label{sect_obs}

As part of our ongoing survey of M101 for massive stars we had already downloaded archival {\it Hubble Space Telescope} (HST) Advanced Camera for Surveys (ACS) images taken with its F435W, F555W and F814W filters to combine with our HST Wide Field Camera 3 (WFC3) data taken with a narrowband F469N filter, tuned to the strong HeII emission line at 468.6 nm. The images were reduced using the \textsc{multidrizzle} routine and mapped onto our WFC3/F469N narrow-band imaging as described in \citet{Shara2013}. Photometry was performed using the standalone \textsc{daophot} code \citep{Stetson1987} and a model point-spread function (PSF) based on isolated point-like stars was built and applied to all other stars in the field. Zero-points from the HST literature were applied to convert our observed magnitudes into Vega magnitudes using the ACS Zeropoint Calculator webtool \footnote[1]{The Zeropoint calculator is available at https://www.stsci.edu/hst/instrumentation/acs/data-analysis/zeropoints} 

\textbf{Using the methods outlined in \citet{Shara2013} we determine our} photometric detection limit for the F435W image to be m$_{F435W}$\,=\,28.5\,mag,  \textbf{but although this is the faintest object we can detect we do not always do so with 100\% efficiency.} \textbf{The magnitude at which we are confident we will always detect an object}, our 100$\%$ detection limit, is m$_{F435W}$\,=\,26.7\,mag which. Using the extinction law of \citet{Cardelli1989} \textbf{and assuming an average extinction of A(H$\alpha$)\,=\,1.06\,mag and a Milky Way foreground reddening of E(B--V)\,=\,0.01\,mag \citep{Lee2009}, corresponding to} A$_{F435W}$\,=\,1.74 mag, \textbf{we determine our absolute magnitude 100\% detection limit} to be M$_{F435W}$\,=\,-–4.07\,mag. The mean Galactic extinction curve does not work well for other galaxies, especially with high star formation rates like M101 \citep{Calzetti1994} so we use values from \citet{Pang2016} for A$_{F555W}$ and A$_{F814W}$ which are presented along with completeness limits for each filter in Table \ref{detection_limits}. \\

\begin{table} %
\centering
	\caption{Archival HST ACS/WFC data used for progenitor identification with the detection limits given in Vega magnitudes. All data were obtained under program ID 9490 (PI: Kuntz). }
    \begin{tabular}{ccccccc}
    	\hline
    	\hline
 Filter & Date Obs & Exposure & Faintest  & A$_{\lambda}$ & \multicolumn{2}{c}{100$\%$ detection limit} \\
       & & Time (sec)       &    Limit   &   & Apparent & Absolute \\
     \hline 
             F435W & 2002-11-15 & 1620 & 28.5 & 1.74 & 26.7 & --4.07 \\
             F555W & 2002-11-16 & 1440 & 28.0 & 1.36 & 26.6 & --3.79 \\
             F814W & 2002-11-16 & 1440 & 27.3 & 0.80 & 25.8 & --4.03 \\
    	\hline
    \end{tabular}%
	  	\label{detection_limits}
\end{table}%

\textbf{Whilst average extinction values are useful to determine our detection limit for objects in M101, more local values of extinction are required for analysis of the progenitor region and ultimately the determination of the initial mass. The SN site lies $\sim$1\,arcsecond from the H\textsc{II} region NGC\,5461 which has undergone analysis by \citet{Kennicutt1996} who analysed spectra of regions 1105, 1098, 1086 and 1052 shown in their Figure 2b. In Table \ref{HII_regions} we present C(H$\beta$) and R$_{23}$ values from Table 2. in \citet{Kennicutt1996} and determine A$_{F814W}$\,=\,0.49$^{+0.359}_{--0.158}$. We note that the R$_{23}$ indicator is double--valued and thus we use the calibration from equation 8 in \citet{Yin2007} to calculate the lower branch 12$+$log(O/H) and \citet{Pilyugin2005} with an excitation parameter P=0.9 for the upper branch. Metallicities derived from these equations for each of the four H\,\textsc{II} regions local to the SN site are presented in Table \ref{HII_regions}. The average metallicity for the lower branch is 12$+$log(O/H)\,=\,7.485 and 12$+$log(O/H)\,=\,8.583 for the upper branch. We note that no H$\alpha$ fluxes are presented in \citet{Kennicutt1996} so the O3N2 or N2 calibrators from e.g. \citet{Pettini2004} cannot be used. However, metallicity determinations of other nearby H\,\textsc{II} regions, albeit further from the SN site, presented in \citet{Pledger2018} using the O3N2 calibration suggests the upper branch is more likely.}

\begin{table}[]
    \centering
       \caption{Four H\textsc{II} regions close to the site of SN2023ixf. C(H$\beta$) and R$_{23}$ are taken from \citet{Kennicutt1996} and we note ID names correspond to those in \citet{Hodge1990}.}
    \begin{tabular}{ccccccc}
     \hline
    \hline
        ID & C(H$\beta$) & E(B-V) &  A$_{F814W}$ & R$_{23}$ & 12$+$log(O/H)$_{lower}$ & 12$+$log(O/H)$_{upper}$  \\
        \hline
        1105 & 0.36 & 0.277 & 0.413 & 6.01 & 7.58 & 8.53\\
        1098 & 0.74 & 0.570 & 0.849 & 4.50 & 7.40 & 8.63 \\
        1086 & 0.32 & 0.246 & 0.362 & 5.02 & 7.47 & 8.59 \\
        1052 & 0.29 & 0.223 & 0.332 & 5.22 & 7.49 & 8.58\\
        \hline
        Average & 0.428 & 0.329 & 0.490 & 5.188 & 7.485 & 8.583 \\
    \end{tabular}
   
    \label{HII_regions}
\end{table}

\section{Identification of the SN progenitor}   \label{sect_res}

\begin{figure}
    \centering
    \includegraphics[width=0.46\columnwidth, trim={0 0 0 0},clip]{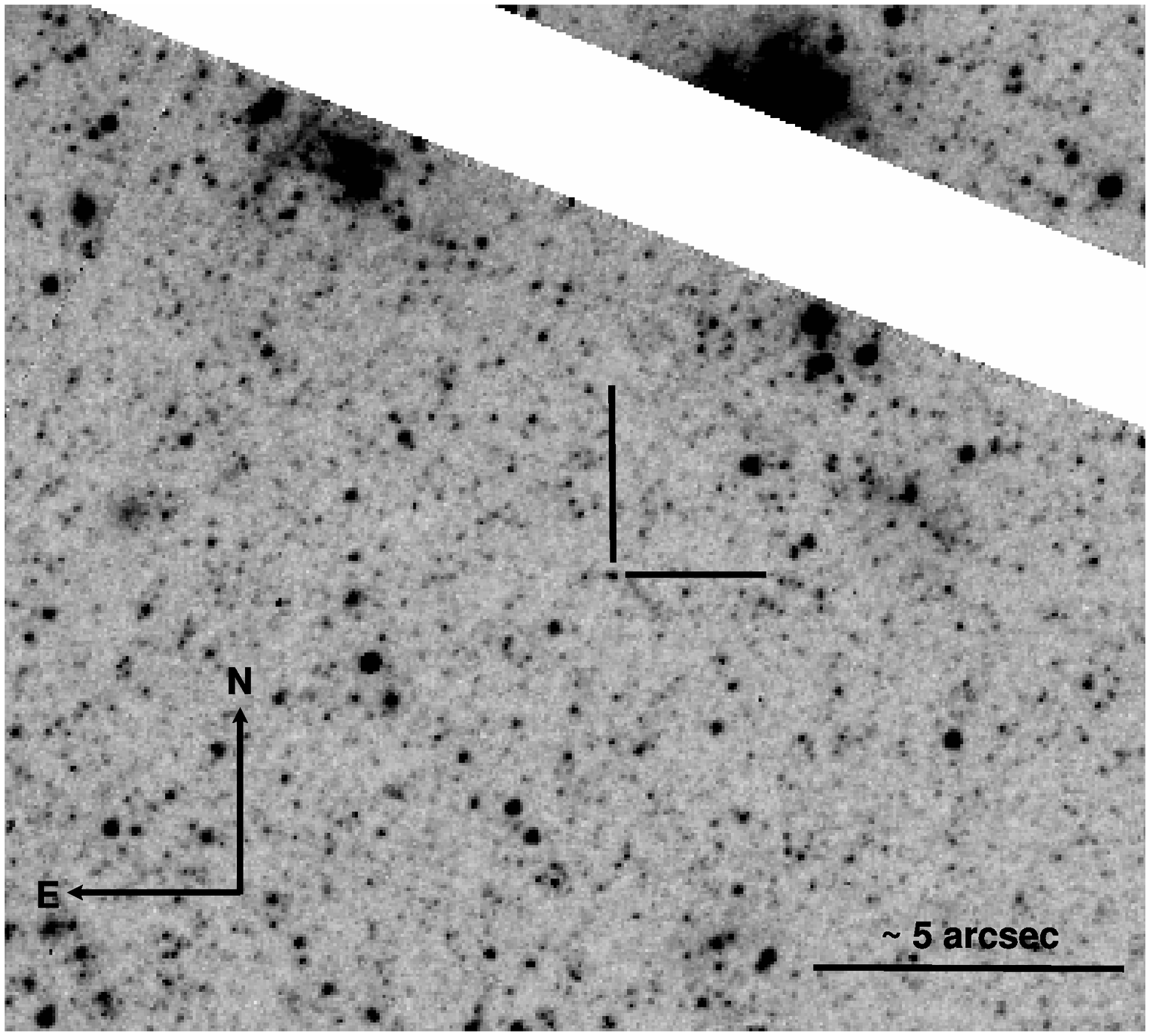}
    \includegraphics[width=0.46\columnwidth, ]{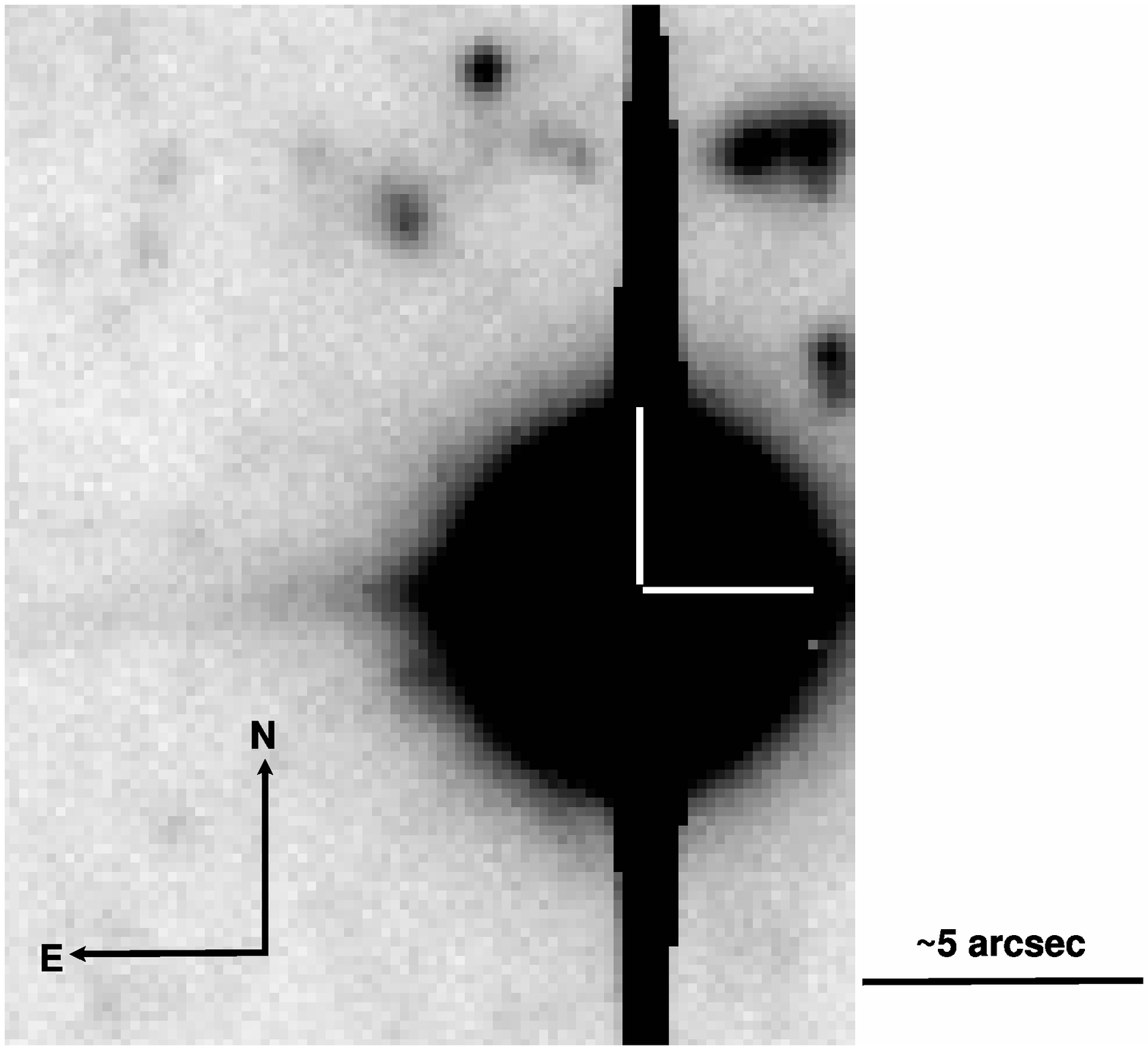}
    \caption{HST/ACS F814W image taken in November 2002 (left) and Gemini/GMOS r-band image taken in June 2023 (right) at the site of SN2023ixf in M101 which became available after the initial submission of this paper. The location of the SN is highlighted and is co-incident with a single source in the F814W image. We note the SN is on the edge of one of the GMOS-N chips.}
    \label{gemini}
\end{figure}

\textbf{To identify any potential progenitors of SN2023ixf we must compare the location of the SN with pre-SN imaging which in turn requires a consistent co-ordinate system between the images. From our previous work our HST images in the three different filters have already been aligned using a geometric transformation from the \textsc{GEOMAP} routine within the HST \textsc{MULTIDRIZZLE} software \citep{Shara2013}.  We applied the same method to r--band Gemini/GMOS-N observations of SN2023ixf (PI: Lotz, Program ID: GN-2003A-DD-105) taken on 05 June 2023. We compared the positions of 30 sources common to both images which yielded a geometric transformation of the Gemini image onto the F814W/ACS image with an RMS error of $\pm$24mas. We applied this transformation using \textsc{GEOTRAN}, and then compared co-ordinates of our targets again, finding a standard deviation of $\sigma$\,=\,0.03\,arseconds. This allows us to determine an accurate position for the SN in the archival imaging as shown in Figure \ref{gemini}.}

\begin{figure}
    \includegraphics[width=0.5\columnwidth, angle=0]{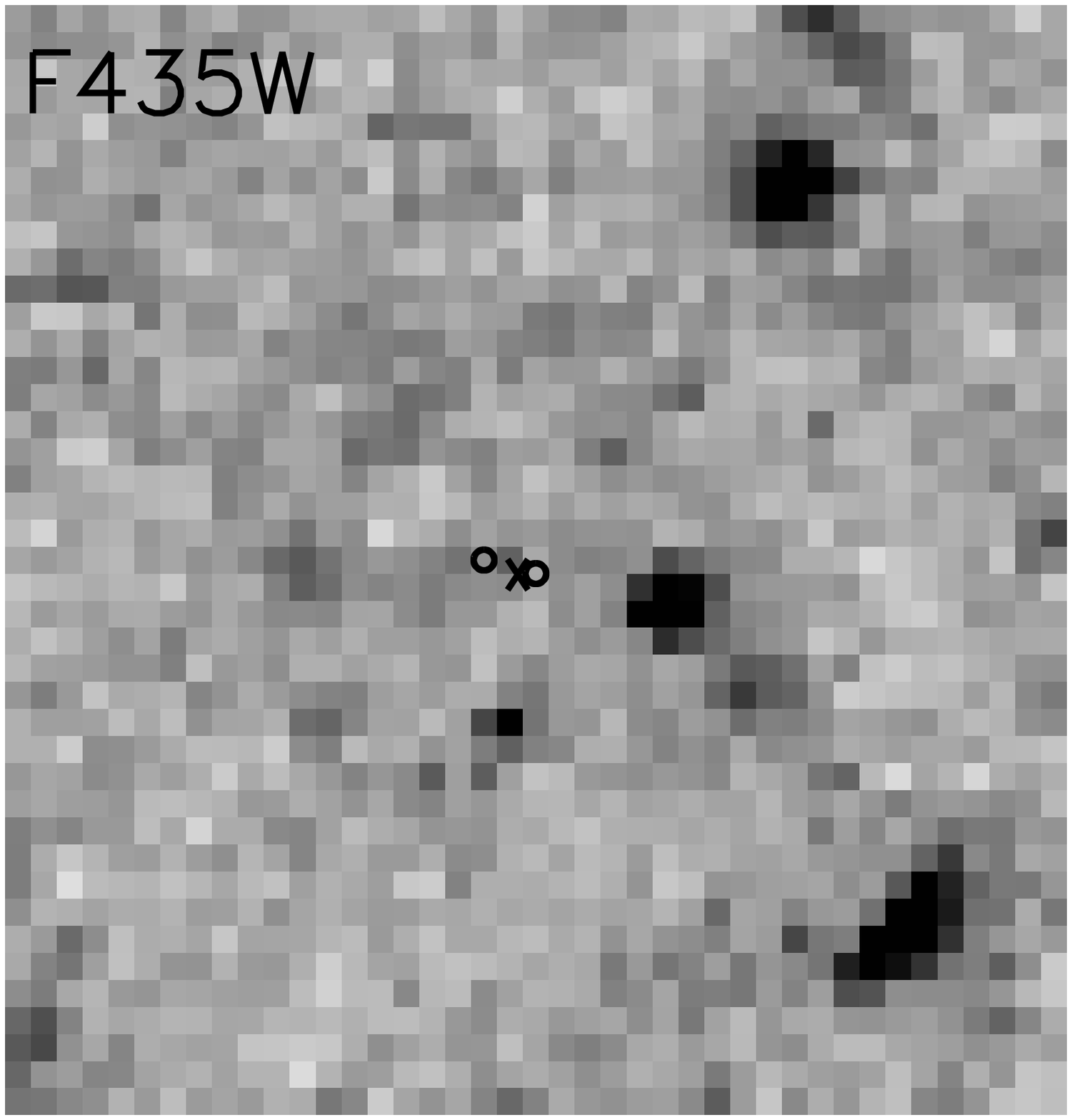}
    \includegraphics[width=0.5\columnwidth, angle=0]{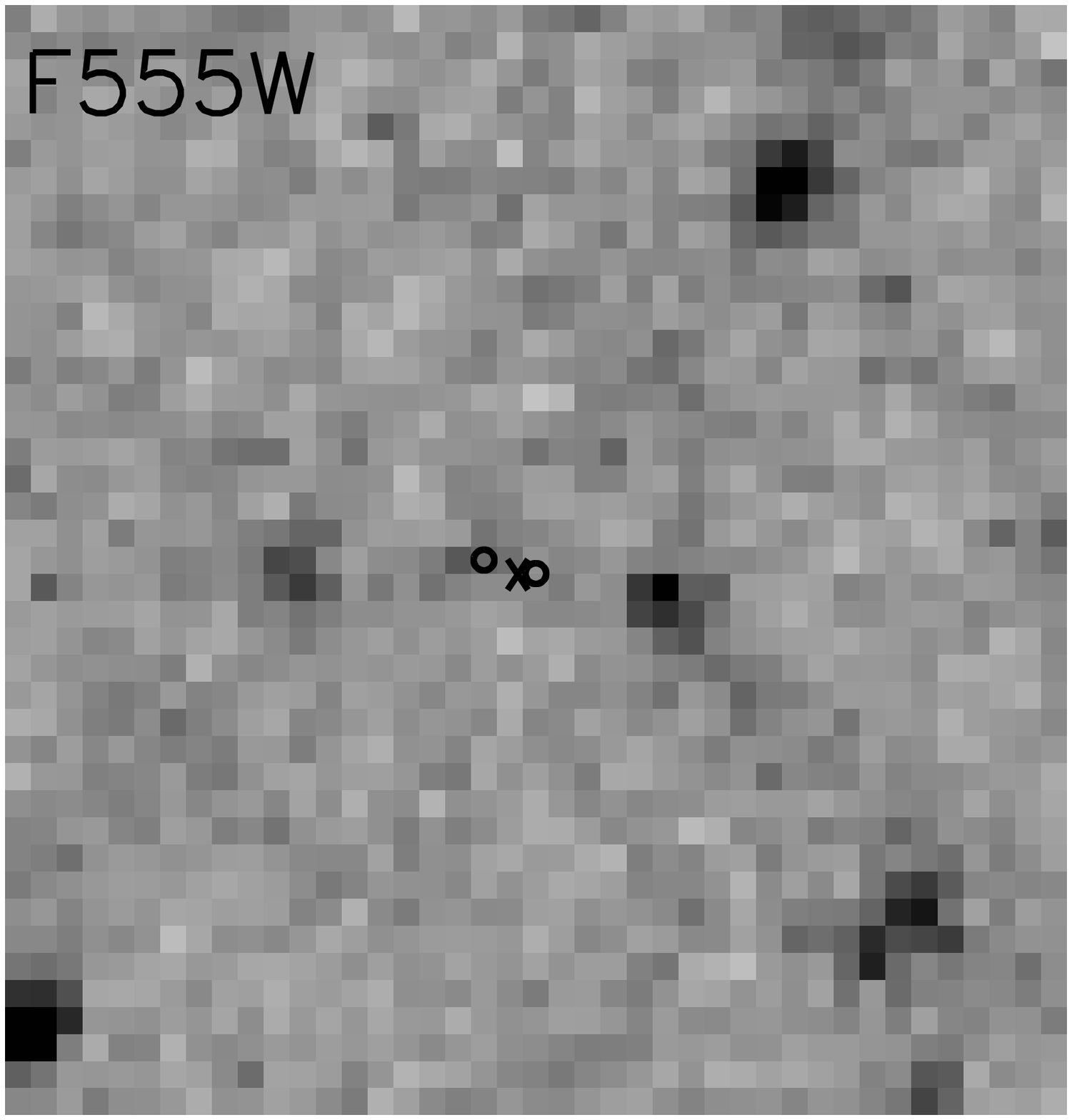}
    \includegraphics[width=0.5\columnwidth, angle=0]{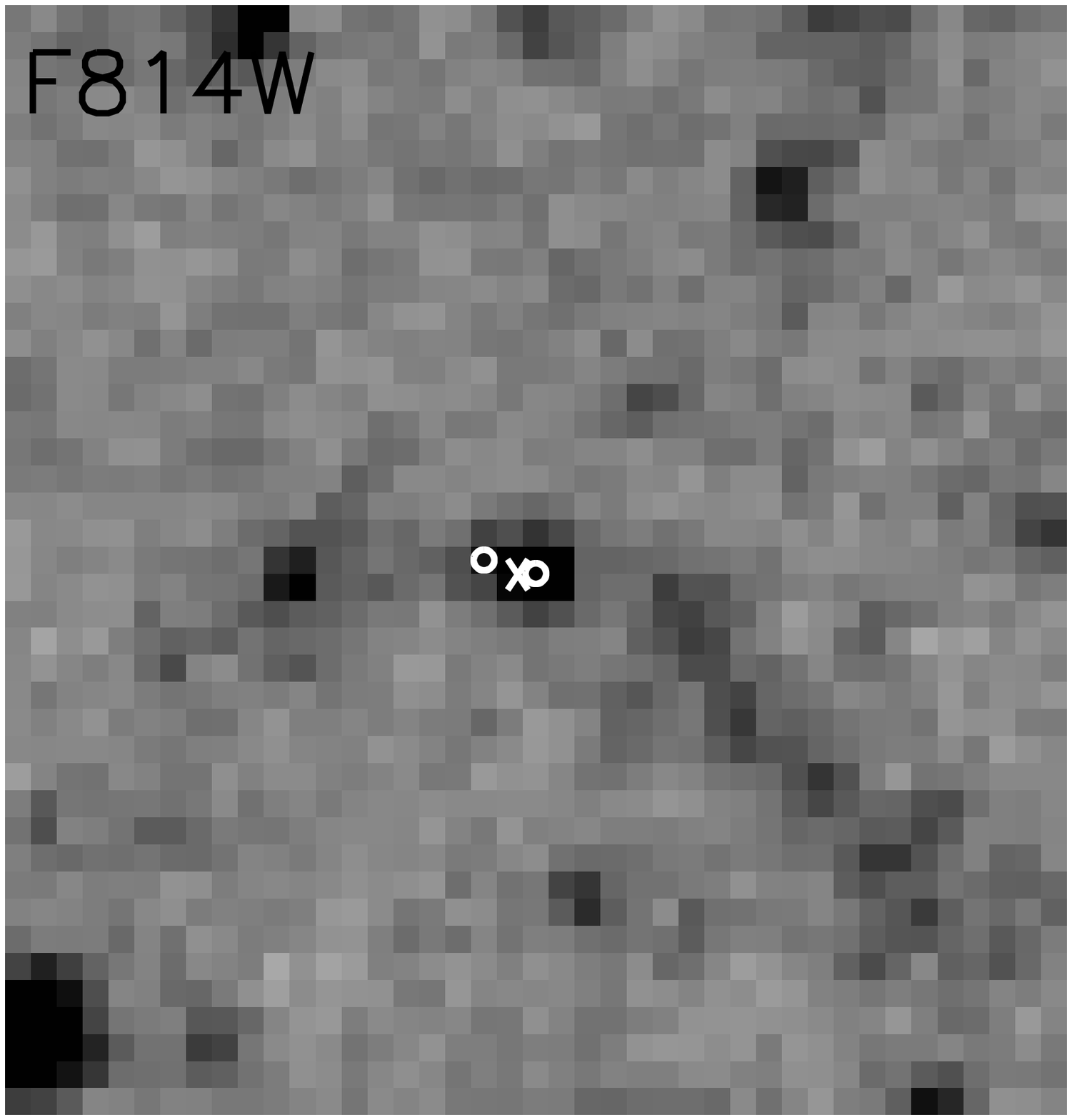}
 \caption{Hubble Space Telescope ACS/WFC images of the $\sim$2 arcsecond$^{2}$ region around the site of SN2023ixf in F435W (top left), F555W (top right) and F814W (bottom) filters. The location of the two sources identified in the F814W image are marked with a circle (with the size corresponding to 1$\sigma$ positional error) whilst the SN location is indicated by an X. We note that North is up and East is left.}
\label{HST_images}
\end{figure}

Using this position and error we detected no progenitor in the F435W image, with the nearest source $\sim$0.44\,arcseconds to the south-west of the supernova location. The F555W image also shows no obvious photometric detection, though one could argue at a hint of a source at the detection limit but slightly off centred from the SN location as shown in Figure \ref{HST_images}. Unfortunately our WFC3/F469N images just miss the SN site, however a Type II progenitor would not be expected to be a star with strong helium emission such as a Wolf Rayet star.

\begin{figure}
\centering
\includegraphics[width=0.48\columnwidth]{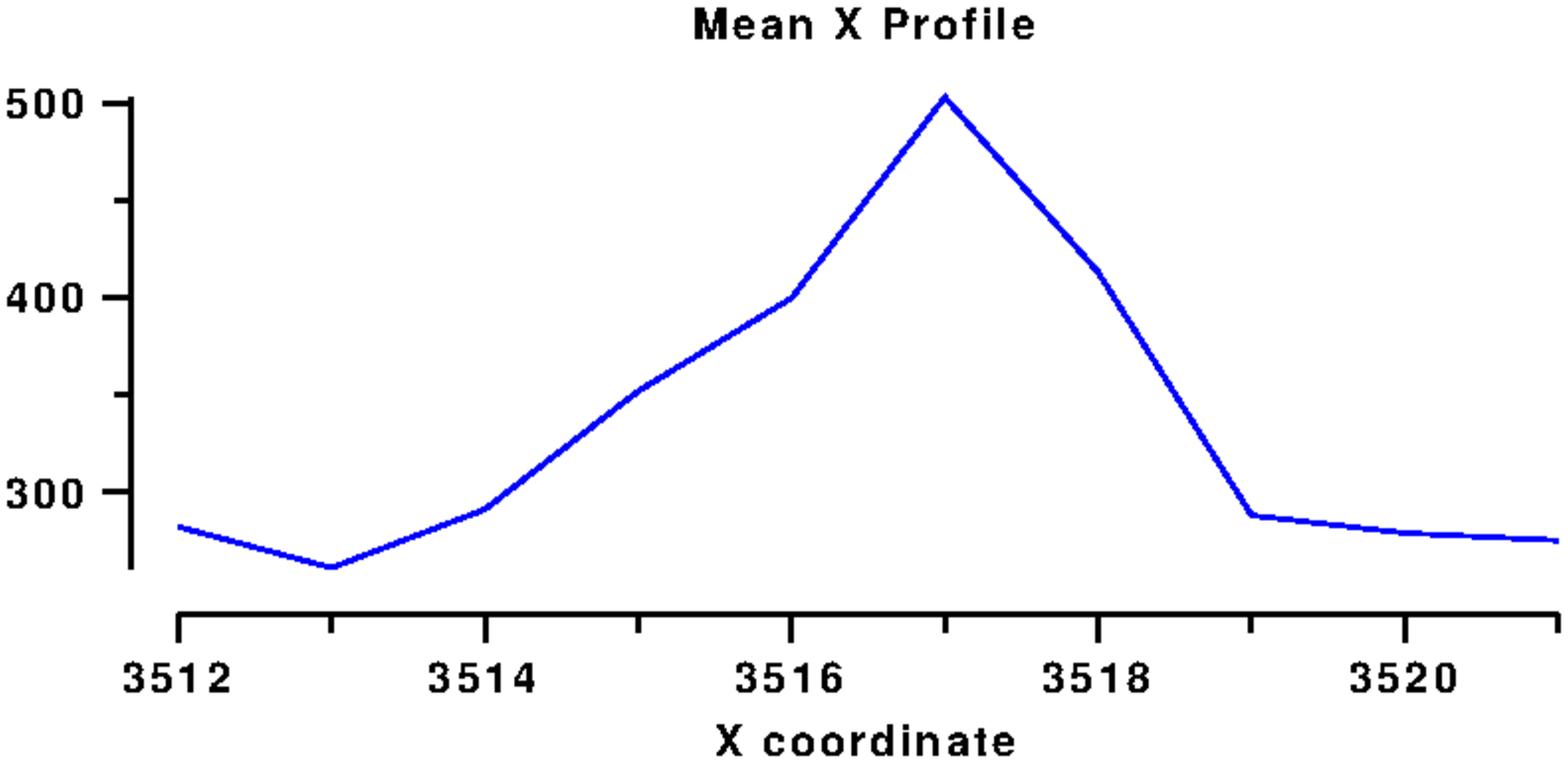}
\includegraphics[width=0.48\columnwidth]{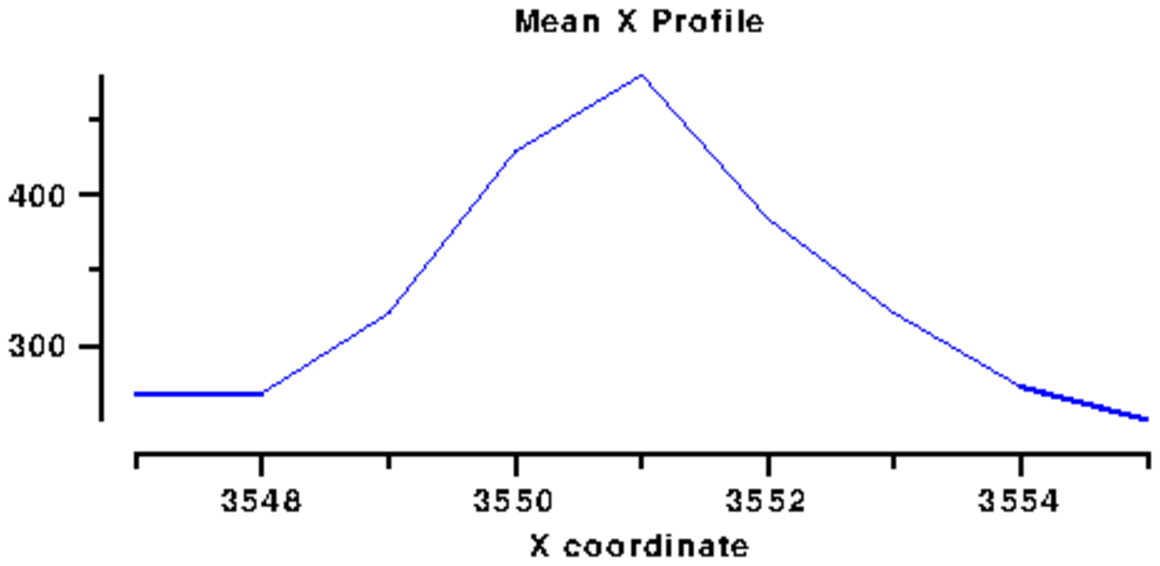}
\caption{Mean X-profile of the source at the location of SN2023ixf (left). The peak at x$\sim$3517 corresponds to the brightest part of the source and another peak is suggested at x$\sim$3515 for the second, fainter source. For context the profile of a nearby point-like star is shown (right) for comparison.}
\label{profile}
\end{figure}

The F814W image reveals a source co-incident with the SN location, within our positional error limit (Figure \ref{HST_images}). Inspection of the profile of the source reveals an asymmetric profile as shown in Figure \ref{profile} suggesting that the source is two partially resolved stars which would be consistent with the hint of an offset source in the F555W image. \textbf{This is supported further from comparison with a nearby point-like star which does not show any evidence of an asymmetric profile.} At a distance of 6.4\,Mpc, the 0.15\,arcsecond spatial resolution of our drizzled HST/ACS images corresponds to a physical size of 4.65\,pc which could easily hide multiple stars. \textbf{However, our positional error suggests that the more prominent source in F814W is co-incident with the SN site within 1$\sigma$ whereas the fainter, more easterly source is further away.}

 Our photometry of these two partially unresolved sources shows that the westerly source (located at $\alpha$\,=\,14:03:38.544 and $\delta$\,=\,54:18:41.90 ) has an apparent Vega magnitude of m$_{F814W}$\,=\,24.41$\pm$0.06 mag, \textbf{which assuming our average extinction A$_{F814W}$\,=\,0.49$^{+0.359}_{-0.158}$\,mag from Table \ref{HII_regions} and a distance of 6.4$\pm$0.7\,Mpc \citep{Shappee2011}, corresponds to an absolute magnitude of M$_{F814W}$\,=\,--5.11$^{+0.65}_{-0.47}$\,mag. Following the initial submission of this paper we note that \citet{Soraisam2023} (ATel $\#16050$) find an apparent magnitude of m$_{F814W}$\,=\,24.39$\pm$0.08\,mag consistent with ours and that \citet{Jacobson2023} find E(B-V)\,=\,0.033\,mag for SN2023ixf based on optical spectra and fitting of the Na\textsc{I} D absorption line which would be most consistent with the lower end of our absolute magnitude.} The other possible progenitor is the eastern source (located at $\alpha$\,=\,14:03:38.552 and $\delta$\,=\,54:18:41.90). It is not well-enough resolved for a magnitude to be determined. \\

\begin{figure}
    \centering
    \includegraphics[width=0.7\columnwidth, angle=-90]{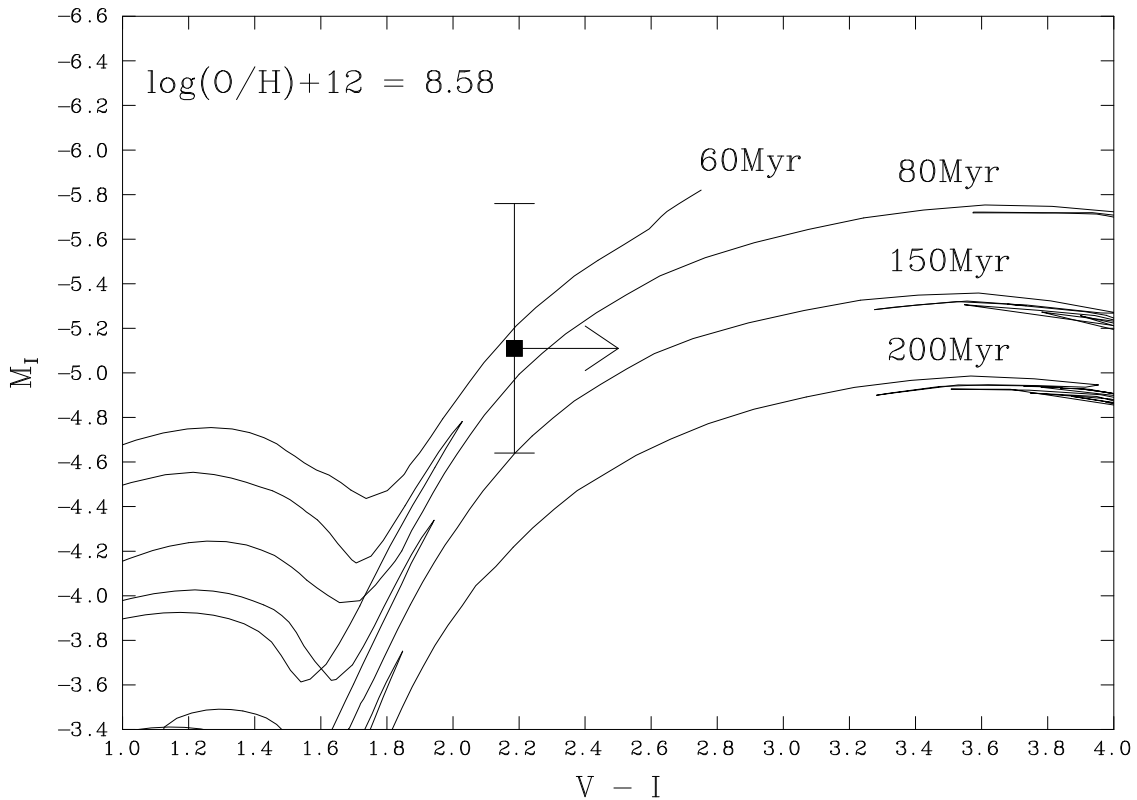}
    \caption{Color-magnitude diagram plotting 80-200Myr PARSEC stellar evolutionary isochrones along with our progenitor candidate at a metallicity of log(O/H)$+$12\,=\,8.58. The V-I plotted is a lower limit using the 100\% detection limit of our F555W images and as such we expect the true value to be further to the right on this plot. We also note that the star will should explode at the end of the evolutionary track.}
    \label{Isochrones}
\end{figure}

\section{Mass Estimate}

To determine a mass estimate for the progenitor we used \textsc{PARSEC}\footnote[2]{PARSEC isochrones are freely available from http://stev.oapd.inaf.it/} stellar evolutionary isochrones with ages 50–200\,Myr from \citet{Bressan2012}. \textbf{We ran models based on both our upper and lower metallicity values using our average extinction value of A$_{F814W}$\,=\,0.49\,mag and use our absolute F814W magnitude of --5.11\,mag, assuming a lower limit of V -- I\,$\sim$\,2.2\,mag based on our 100\% detection limit in F555W of 26.6\,mag to assess the initial mass of the progenitor. Error bars shown in Figure \ref{Isochrones} are based on errors in distance, extinction and magnitude. Figure \ref{Isochrones} show the stellar evolutionary tracks that best match our observations assuming a metallicity of log(O/H)$+$12\,=\,8.58. Within our error bars the progenitor star is $\sim$60-200\,Myr old which corresponds to an initial mass of $\sim$4--7M\,$_{\odot}$ given the star would explode at the end of the track. We note that stars in the lower end of this mass range would not be expected to explode as a RSG. Similarly, if we assume the lower metallicity of log(O/H)$+$12\,=\,7.49 then the luminosity falls below that expected to explode as a RSG, consistent with a comparison with BPASS models (J. Eldridge; private communication; \cite{Stanway2018}). We propose that the likely explanation for this low mass is that there is a lot of dust that we don't account for and that the progenitor's true initial mass is in the lower mass range for RSG of 8-10\,M$_{\odot}$. The presence of dust is consistent with the findings of \citet{Szalai2023}(ATel $\#$16042) who suggest a mass of $\sim$15M$_{\odot}$ from Spitzer data.  Further analysis of the model-dependent dust, combined with light curve modelling, is required to fully assess the nature of the progenitor.}

\section{Conclusions}  \label{sect_dis}

We have detected a possible optical counterpart to the progenitor of the type IIP supernova 2023ixf in archival {\it Hubble Space Telescope} images. The counterpart is not seen in (blue) F435W or (visible) F555W filter images, but is easily detected in (NIR) F814W images. The red color and brightness of the progenitor are consistent with it being a red supergiant of at least 7M$_{\odot}$, likely higher at $\sim$8--10$_{\odot}$ if there is significant dust present. There is also an extremely faint, unresolved source that could also potentially be the progenitor of SN2023ixf although our astrometric calibration suggests this is unlikely. There is a suggestion of a detection in the F555W image offset from the F814W source but this would suggest a bluer object not typical of Type II SN progenitors. Post-SN imaging with HST will be crucial to confirm the disappearance of the progenitor, as is modelling of the SN light-curve and the dust surrounding the supernova.

\
\section{Acknowledgments}

MMS thanks James Garland for alerting him to this nearby supernova.
We thank David Zurek for advice on magnitude system conversions and Anne Sansom for advice on stellar isochrones. We also thank Jan Eldridge for discussion of the progenitor mass in the BPASS models.
Some of the data presented in this paper were obtained from the Mikulski Archive for Space Telescopes (MAST) at the Space Telescope Science Institute. The data used from HST is available at MAST: \dataset[10.17909/qbmp-nw13]{https://doi.org/10.17909/qbmp-nw13}. Some of this work was supported by the generous support of the late Hilary and Ethel Lipsitz and we gratefully acknowledge them for their support. This research is based on NASA/ESA Hubble Space Telescope observations obtained at the Space Telescope Science Institute, which is operated by the Association of Universities for Research in Astronomy Inc. under NASA contract NAS5-26555. Based on observations obtained at the international Gemini Observatory, a program of NSF’s NOIRLab, which is managed by the Association of Universities for Research in Astronomy (AURA) under a cooperative agreement with the National Science Foundation on behalf of the Gemini Observatory partnership: the National Science Foundation (United States), National Research Council (Canada), Agencia Nacional de Investigaci\'{o}n y Desarrollo (Chile), Ministerio de Ciencia, Tecnolog\'{i}a e Innovaci\'{o}n (Argentina), Minist\'{e}rio da Ci\^{e}ncia, Tecnologia, Inova\c{c}\~{o}es e Comunica\c{c}\~{o}es (Brazil), and Korea Astronomy and Space Science Institute (Republic of Korea).

%

\vspace{5mm}
\facilities{HST(ACS/WFC3), Gemini(GMOS-N)}


\software{{\sc{MULTIDRIZZLE}} (https://www.stsci.edu/~koekemoe/multidrizzle/),
\textsc{IRAF} \citep{Tody1986}, \textsc{PARSEC} isochrones are freely available from http://stev.oapd.inaf.it/}



\bibliography{SNref}{}
\bibliographystyle{aasjournal}

\end{document}